\documentclass[%
 reprint,
 nofootinbib,
 amsmath,amssymb,
 aps,
]{revtex4-2}

\usepackage{graphicx}%
\usepackage{subcaption}%
\usepackage{dcolumn}%
\usepackage{bm}%
\usepackage{bbold} %
\usepackage{hyperref}%
\usepackage{siunitx}%
\usepackage{physics}

\begin{document}

\renewcommand{\tableautorefname}{Tab.}
\renewcommand{\figureautorefname}{Fig.}
\renewcommand{\equationautorefname}{Eq.}

\preprint{}

\title{Post-hoc regularisation of unfolded cross-section measurements}

\author{Lukas Koch}
\email{lukas.koch@uni-mainz.de}
\altaffiliation[previously at ]{%
University of Oxford,
Particle Physics Department,
DWB, Keble Road,
OX1 3RH, Oxford,
United Kingdom
}%
\affiliation{%
Johannes Gutenberg University Mainz\\
Institute of Physics\\
Staudingerweg 7\\
55128 Mainz, Germany
}%

\date{\today}%

\begin{abstract}
Neutrino cross-section measurements are often presented as unfolded binned distributions in ``true'' variables.
The ill-posedness of the unfolding problem can lead to results with strong anti-correlations and fluctuations between bins, which make comparisons to theoretical models in plots difficult.
To alleviate this problem, one can introduce regularisation terms in the unfolding procedure.
These suppress the anti-correlations in the result, at the cost of introducing some bias towards the expected shape of the data.

This paper discusses a method using simple linear algebra, which makes it is possible to regularise any result that is presented as a central value and a covariance matrix.
This ``post-hoc'' regularisation is generally much faster than repeating the unfolding method with different regularisation terms.
The method also yields a regularisation matrix which connects the regularised to the unregularised result,
and can be used to retain the full statistical power of the unregularised result when publishing a nicer looking regularised result.

In addition to the regularisation method, this paper also presents some thoughts on the presentation of correlated data in general.
When using the proposed method, the bias of the regularisation can be understood as a data visualisation problem rather than a statistical one.
The strength of the regularisation can be chosen by minimising the difference between the implicitly uncorrelated distribution shown in the plots and the actual distribution described by the unregularised central value and covariance.

Aside from minimising the difference between the shown and the actual result, additional information can be provided by showing the local log-likelihood gradient of the models shown in the plots.
This adds more information about where the model is ``pulled'' by the data than just comparing the bin values to the data's central values.
\end{abstract}

\maketitle

\section{Introduction}

Neutrino cross-section measurements are often presented as unfolded binned distributions in ``true'' variables.
But unfolding can lead to results with large bin-to-bin variations and strongly correlated data points.
These results are statistically ``correct'' in the sense that they have the minimum amount of bias and the right coverage properties \cite{Cowan1998}.
But they are ``ugly'' when plotted and can be difficult to interpret beyond a global goodness of fit (GOF).
To reduce these problems, one can add regularisation to the unfolding procedure.

This is usually done via some penalty term that pulls the result towards some expected property, like ``smoothness'' of bin-to-bin differences.
Regularised results show a smaller variances and less bin-to-bin correlations,
but they are necessarily biased towards the aim of the regularisation.
To provide readable plots as well as maximally ``correct'' data for statistical tests, some experiments report both a regularised as well as an unregularised version of their results, e.g. \cite{T2K:2018rnz,T2K:2020jav}.

Another way of controlling bin variation and correlations is by choosing a wider binning for the result.
This can make the value of the single bins better constrained, but it reduces the power of the data,
as one reduces the resolution of the result.
Depending on how this is done, it could also lead to increased model dependencies
if the detector response varies considerably within a single analysis bin.
The latter can be avoided by unfolding with fine bins and then combining them into larger bins for presentation \cite{Stanley2021}.
In that sense, the re-binning is a form of regularisation, where joined cross-section bins are forced to have identical values.

Tang et al presented an unfolding and regularisation scheme based on a Wiener filter \cite{Tang2017},
which is used in recent MicroBooNE papers \cite{Abratenko2021}.
The data is unfolded without regularisation using the (pseudo-)inverse of the smearing matrix $R$.
Then the unregularised result is multiplied with an ``additional smearing matrix'' $A$,
which provides the regularisation.
The matrix $A$ is constructed using the smearing matrix, the covariance matrix of the unregularisd result, and the expected true spectrum (i.e. the nominal model).

There are two ways of interpreting the use of the matrix $A$.
It can either be seen as a coordinate transformation, or as a transformation of the result.
In the former case, the result itself, i.e. what it says about reality, is not changed,
but the meaning of the variables, i.e. the bin values, \emph{does} change.
It is like expressing the result in a coordinate system with rotated axes.
In the second interpretation, the meaning of the axes is kept, but the result itself is changed.
I.e. the likelihood of a point in the parameter space is different when calculated with the regularised result compared to the unregularised one.

By publishing the regularised result together with the matrix $A$, the publication contains exactly the same information as the unregularised result.
I.e. by ``smearing'' some model expectation with $A$ before calculating the Mahalanobis distance (M-distance) \footnote{The squared M-distance is often called ``the chi-square`` in physics.},
one will recover exactly the same values as if the comparison happened directly with the unregularised result.
This means that all the desirable properties of the unregularised result -- minimal bias and correct coverage properties -- are still present.
This makes sense when viewing $A$ not as a smearing matrix, but as a coordinate transformation.
A multivariate normal distribution will still be a multivariate normal distribution after a linear coordinate transformation.
No information is lost, provided the matrix $A$ is a full-rank matrix.

Compared to the practice of providing the regularised and unregularised results separately,
this approach has the advantage of being completely self-consistent.
There is a direct link between regularised and unregularised result: the matrix $A$.
So there is less room for confusion about which is the ``real'' result, or which result should be shown in plots or used for quantitative model comparisons and fits.

Note that the matrix $A$ does not need to conserve the total number of events, nor do the elements need to be positive.
It is a more general transformation than a classical smearing matrix.
That is why we will use the term ``regularisation matrix'' in this paper instead.

\section{Regularised likelihood fits}

One quite common method of unfolding is the minimisation of the negative log-likelihood function:
\begin{equation}
    -2 \ln L(\bm\theta) = -2 \ln L_\text{stat}(\bm\theta) -2 \ln L_\text{syst}(\bm\theta) \text{,}
\end{equation}
where $\bm\theta$ is a vector of parameters, which includes the cross-section parameters of interest as well as nuisance parameters.
$L_\text{stat}$ describes the statistical likelihood of the observed data, e.g. a Poisson likelihood,
while $L_\text{syst}$ describes external constraints on the nuisance parameters.

The output of a fit like this is the maximum likelihood estimator (MLE) $\bm{\hat\theta}$ as well as a covariance matrix $V$ that describes the likelihood surface around that best fit point:
\begin{equation}\label{eq:approx}
    -2 \ln L(\bm\theta) \approx \qty(\bm\theta - \bm{\hat\theta})^T V^{-1} \qty(\bm\theta - \bm{\hat\theta}) + \text{const.}
\end{equation}
These can then be used to construct confidence intervals in the usual manner.

This explicit likelihood maximisation is as susceptible to statistical fluctuations as the implicit likelihood maximisation when simply inverting the detector smearing matrix.
So to regularise the result, an additional penalty term $P_\text{reg}$ is added to the likelihood function:
\begin{equation}
    -2 \ln L'(\bm\theta) = -2 \ln L(\bm\theta) + P_\text{reg}(\bm\theta) \text{.}
\end{equation}
Conceptually, this adds an external constraint on the parameters of interest,
e.g. the expectation that the difference between neighbouring bins in a differential measurement should be smooth rather than wildly fluctuating up and down.
If the regularisation term can be expressed with a symmetric penalty matrix $Q$, it is called ``Tikhonov regularistaion'' or ``ridge regression'' \cite{Tikhonov1977,Gruber1998}\footnote{The penalty term is often expressed as $||C\bm\theta||^2_2$, which is equivalent to the convention used in this paper when $Q = C^TC$.}:
\begin{equation}\label{eq:tikhonov}
    -2 \ln L'(\bm\theta) = -2 \ln L(\bm\theta) + \bm\theta^T Q \bm\theta \text{.}
\end{equation}
This yields a new MLE $\bm{\hat\theta}'$ as well as a new covariance matrix around that point $V^*$.

Using \autoref{eq:approx} we can express this as
\begin{align}
    -2 \ln L'(\bm\theta) &\approx \qty(\bm\theta - \bm{\hat\theta})^T V^{-1} \qty(\bm\theta - \bm{\hat\theta}) + \bm\theta^T Q \bm\theta + \text{const.} \\
    &= \qty(\bm\theta - \bm{\hat\theta}')^T V^{*-1} \qty(\bm\theta - \bm{\hat\theta}') + \text{const.}
\end{align}
with
\begin{gather}
    V^* = (V^{-1} + Q)^{-1} \text{,}\nonumber\\
    \bm{\hat\theta}' = A \bm{\hat\theta} \text{,}\\
    V' = AVA^T \text{,}\\
    A = \qty(V^{-1} + Q)^{-1}V^{-1} \text{.}
\end{gather}
Here $V^*$ is the covariance matrix of the regularised likelihood surface.
The maximum regularised likelihood estimator $\bm{\hat\theta}'$, is a linear transformation of the original best-fit point $\bm{\hat\theta}$,
so it has the variance $V'$, which we get from linear error propagation.
This variance is different from $V^*$, because the regularisation is constant and not a randomly shifted likelihood coming from a second measurement.
This approach requires that the new MLE is close enough to the original one so the quadratic parameterisation of the original likelihood surface is sufficiently accurate.

What this means is that we do not have to re-do the likelihood fit to get a regularised result.
We can simply calculate $A$ from the unregularised covariance $V$ and the penalty matrix $Q$, and then apply it to the unregularised result.
Not only does this allow us to publish a single regularised result with the full unregularised  coverage properties (using the matrix $A$).
In general it is also much faster than re-running the fit.
This makes it much easier to do studies with different $Q$ matrices,
e.g. to choose a strength of the regularisation that balances the increased bias with the reduced variation of the regularised result.

Note that this approach does not depend on the details of the fitting procedure or the detector response.
It can be applied to any measurement that parameterises the likelihood function around the MLE with a covariance matrix,
or even Bayesian results that report the posterior probability as a multivariate normal distribution.
In that sense, it is a data driven method with the only model assumptions going into the choice of $Q$.
It is thus possible to apply this method to published results even long after the original analysis code has been lost and details about the detector response have passed into legend.
When used like this, this method could be called ``post-hoc regularisation''.

Care needs to be taken when this method is applied to results that have already been regularised, though.
Applying additional regularisation would be equivalent to adding additional penalty terms to the likelihood function.
E.g. if the original regularisation can be expressed as in \autoref{eq:tikhonov}, it would mean using an effective penalty matrix $Q$ that is the sum of the original and additional matrices.
This could have unintended consequences regarding what the combined matrix is actually penalising against.
The method presented here can still be employed as a data visualisation technique to present the original regularised result as faithfully as possible in plots of the data (see the following sections),
especially if the original result is only ``lightly regularised'' and still contains considerable (anti-)correlations.
Whenever possible, it should be used with the unregularised data though, as that would ensure the best use of the actual data constraints, and the least amount of bias in the GOF calculations.

\section{Choosing the penalty matrix}

The regularisation matrix $A$ is completely determined by the unregularised covariance $V$ and the penalty matrix $Q$.
The choice of $Q$ is somewhat arbitrary, but we can investigate some common patterns for choosing it.

A simple choice for $Q$ is to penalise differences between neighbouring bins in a differential cross section:
\begin{equation}
    P_\text{reg}(\bm\theta) = \tau \sum_{i=1}^{N-1} (\theta_i - \theta_{i+1})^2 \text{,}
\end{equation}
where $\theta_{1\dots N}$ are the parameters of interest corresponding to the truth-space cross-section bins,
and the free parameter $\tau$ is the ``regularisation strength''.
This can be expressed as a penalty matrix:
\begin{gather}
    P_\text{reg}(\bm\theta) = \bm\theta^T Q \bm\theta \text{,}
\end{gather}
\begin{gather}
    Q = \tau Q_1 = \tau \mqty(
        1 & -1 & 0 & 0 & \\
        -1 & 2 & -1 & 0 & \dots \\
        0 & -1 & 2 & -1 & \\
        0 & 0 & -1 & 2 & \\
         & \vdots &  & & \ddots
        )\text{.}
\end{gather}
This would bias the result towards a flat solution, i.e. all elements of $\bm\theta$ being equal.

Rather than doing that, it is also possible to put a penalty on different scaling compared to a reference model shape $\bm{m}$:
\begin{align}
    P_\text{reg}(\bm\theta) &= \tau \sum_{i=1}^{N-1} \qty(\frac{\theta_i}{m_i} - \frac{\theta_{i+1}}{m_{i+1}})^2 \\
    &= \bm\theta^T Q \bm\theta = \tau \bm\theta_m^T Q_1 \bm\theta_m \text{,}
\end{align}
with $\theta_{mi} = \theta_i / m_i$ and
\begin{widetext}
\begin{gather}
    Q = \tau Q_{1m} = \tau \mqty(
        1/m_1^2 & -1/(m_1 m_2) & 0 & 0 & \\
        -1 / (m_1 m_2) & 2 / m_2^2 & -1 / (m_2 m_3) & 0 & \dots \\
        0 & -1 / (m_2 m_3) & 2 / m_3^2 & -1 / (m_3 m_4) & \\
        0 & 0 & -1 / (m_3 m_4) & 2 / m_4^2 & \\
         & \vdots &  & & \ddots
        )\text{.}
\end{gather}
\end{widetext}
This now biases the result towards the shape of the reference model, which usually is better motivated than just a flat shape.

These are just two examples of possible $Q$~matrices and different ones are easily constructed or can be found in the literature, e.g. matrices that penalise higher order derivatives of the bin contents as found in \cite{Tang2017}.

The regularisation strength determines how strongly the bin-to-bin differences are penalised.
By changing its value, we can tune the amount of regularisation bias we introduce to the result.
This bias is often expressed as the shift of the MLE in terms of squared M-distance:
\begin{equation}
    D^2_\text{M} = (\bm{\hat\theta}' - \bm{\hat\theta})^T V^{-1} (\bm{\hat\theta}' - \bm{\hat\theta})
\end{equation}
Here we use the covariance of the unregularised result.

The regularised result does differ from the unregularised result by more than just the MLE, though.
It also has a different covariance matrix.
One could imagine a case where the regularisation does not move the MLE, but it reduces the size of the uncertainty considerably.
This affects the coverage properties of the result, but is not reflected at all in the M-distance of just the MLE.

To include the effect of a change of covariance, we can use a the Wasserstein metric \cite{Villani2009}.
The Wasserstein distance (W-distance) of order 2 between two random distributions $X$ and $Y$ is defined as the minimum possible expectation value of the $L_2$ distance (i.e. the Euclidean distance) between $X$ and $Y$ over all possible joint distributions of the two:
\begin{equation}
    D^2_W = \inf E[|\bm{X}-\bm{Y}|^2] \text{,}
\end{equation}
where $E[\,]$ denotes the expectation value and the infimum is taken over all possible joint distributions of $\bm{X}$ and $\bm{Y}$ that retain the original marginal distributions of the two.
In computer science it is also know as the earth mover's distance \cite{Levina2001},
as it is a measure for the minimum distance one needs to ``move probability around'' in order to transform $X$ into $Y$.

For the comparison of two multivariate normal distributions with means $\bm\theta$ and $\bm\theta'$, and covariances $V$ and $V'$, it can be expressed in closed form \cite{Olkin1982,Dowson1982}:
\begin{equation}
    D^2_\text{W} = \qty|\bm{\hat\theta}' - \bm{\hat\theta}|^2 + \Tr(V + V' - 2\qty(V^{1/2}V'V^{1/2})^{1/2})
\end{equation}
$D_\text{W}$ has the unit of the variable space the distributions are defined on.
I.e. for two point-like distributions, the term in the trace vanishes and it is equivalent to the Euclidean distance between the points.

This is not useful for judging the regularisation bias, since it ignores the different importance of different directions, which is given by the covariance matrix.
Furthermore, in case the parameter vector $\bm\theta$ contains elements with different units, the Euclidean distance is undefined.
Both issues can be easily fixed by transforming both distributions into a different coordinate space with a whitening transformation of the unregularised result:
\begin{gather}
    V^{-1} = U^TU \text{,}
    \bm\theta \xrightarrow{} \bm{z} = U \bm\theta \text{,} \\
    V \xrightarrow{} \mathbb{1} \text{,} \\
    \bm\theta' \xrightarrow{} \bm{z}' = U \bm\theta' \text{,} \\
    V' \xrightarrow{} UV'U^T \text{,}
\end{gather}
\begin{widetext}
\begin{align}
    D^2_\text{W} &= \qty|\bm{\hat{z}}' - \bm{\hat{z}}|^2 + \Tr(\mathbb{1} + UV'U^T - 2\qty(\mathbb{1}^{1/2}UV'U^T\mathbb{1}^{1/2})^{1/2}) \\
    &= \qty(\bm{\hat\theta}' - \bm{\hat\theta})^TV^{-1}\qty(\bm{\hat\theta}' - \bm{\hat\theta}) + N + \Tr(UV'U^T - 2\qty(UV'U^T)^{1/2}) \text{.}
\end{align}
\end{widetext}
Here $U^TU$ is a Cholesky decomposition of $V^{-1}$ and $N$ is the number of elements in $\bm\theta$.
In this coordinate system, the unregularised covariance matrix is the identity matrix $\mathbb{1}$, and the Euclidean distance corresponds to the M-distance.
In fact, in case the two covariance matrices are identical $V=V'$, this W-distance is identical to the M-distance:
\begin{gather}
    UV'U^T \overset{V=V'}{=} U\qty(U^TU)^{-1}U^T = \mathbb{1} \text{,}\\
    D^2_\text{W} \overset{V=V'}{=} D^2_\text{M} + N + \Tr(-\mathbb{1}) = D^2_\text{M} \text{.}
\end{gather}
This means we can treat the W-distance as an extension of the M-distance, which takes into account both the difference in central values and covariances.

For Frequentist results, it is strictly speaking not correct to interpret the MLE and covariance matrix as a probability distribution.
They actually just describe a quadratic approximation of the negative log likelihood ratio surface around the minimum (MLE).
The mathematical algorithm to calculate the W-distance still works though,
and all we are interested in here is a measure for the difference of the results that includes the difference in covariance.
So it should be fine to just define this measure even if the original meaning of the W-distance does not technically apply.
In any case, the Frequentist likelihood surface should be mostly equivalent to a Bayesian result with a flat prior in the chosen variable parameterisation.

\begin{figure*}
    \centering
    \includegraphics[width=0.9\textwidth]{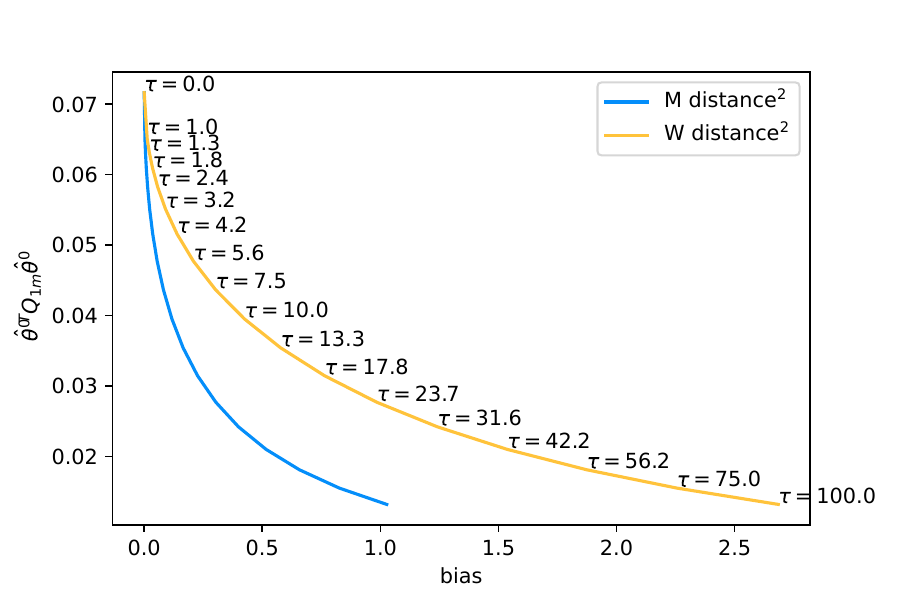}
    \caption{L-curve example plot.
    When increasing the regularisation strength, the value of the penalty term at the MLE decreases,
    while the bias introduced to the result (as measured by the M or W distance to the unregularised result) increases.
    This plot can be used to choose a regularisation strength.
    Points at the lower left corner or slightly above it are common choices.
    }
    \label{fig:L-curve}
\end{figure*}

Now that we have a way of judging the bias introduced by the regularisation,
we also need a measure for the ``gain'' we have from introducing it.
Then we need to balance the two to decide on a regularisation strength.
A common way of doing this is via ``L-curves'', where one plots the penalty term vs the bias for different values of regularisation strength (see \autoref{fig:L-curve}).
The idea is that the penalty term is a measure for the ``undesirableness'' of the result.
After all, the job of the penalty term in the fit is to pull the result towards lower values of the penalty term.
One can then choose a regularisation strength that does reduce this term, while not introducing ``too much'' bias, by picking a point close to the corner point of the curve.

There are a couple of issues with this approach.
First of all, ``a point close to the corner point of the curve'' is not very well defined and the impression of where that point lies will depend on the relative scale of the plot axes.
Secondly, the penalty term is not actually the thing we want to reduce by introducing it to the fit.
It is just a means to the end of making the plots of correlated data points less misleading.
It would thus be desirable to have a more direct measure of the ``badness'' of the plots.

The problem with plots of correlated data points is that the plots generally do not show any of the correlation.
The plotted error bars look exactly the same as error bars for uncorrelated data points.
Our usual ``by eye'' metrics of goodness of fit,
like the expectation that roughly a third of the points should deviate from the model prediction by more than the size of the error bars,
only really apply to uncorrelated data, though.
So since we cannot see the correlations in the plot, and all our intuitions are built on uncorrelated data, we implicitly tend to interpret the plots as if the data were uncorrelated.\footnote{At least the author of this paper does.}

So the ``badness'' of the plot is the difference between the actual correlated likelihood or posterior distribution as described by the central values and covariance matrix, and the uncorrelated distribution as implied by the plot.
And we can quantify this with the W-distance between the two distributions.
For this we just need to set the off-diagonal elements of the regularised covariance matrix to 0 when calculating the W-distance:
\begin{widetext}
\begin{gather}
    D^2_\text{W,plot} = \qty|\bm{\hat\theta}' - \bm{\hat\theta}|^2 + \Tr(V + V'^* - 2\qty(V^{1/2}V'^*V^{1/2})^{1/2}) \text{,}
\end{gather}
\end{widetext}
with $V'^*_{ij} = \delta_{ij}V'_{ij}$, where $\delta_{ij}$ is the Kronecker delta.
We call $D^2_\text{W,plot}$ the ``plot bias''.

\begin{figure*}
    \centering
    \includegraphics[width=0.9\textwidth]{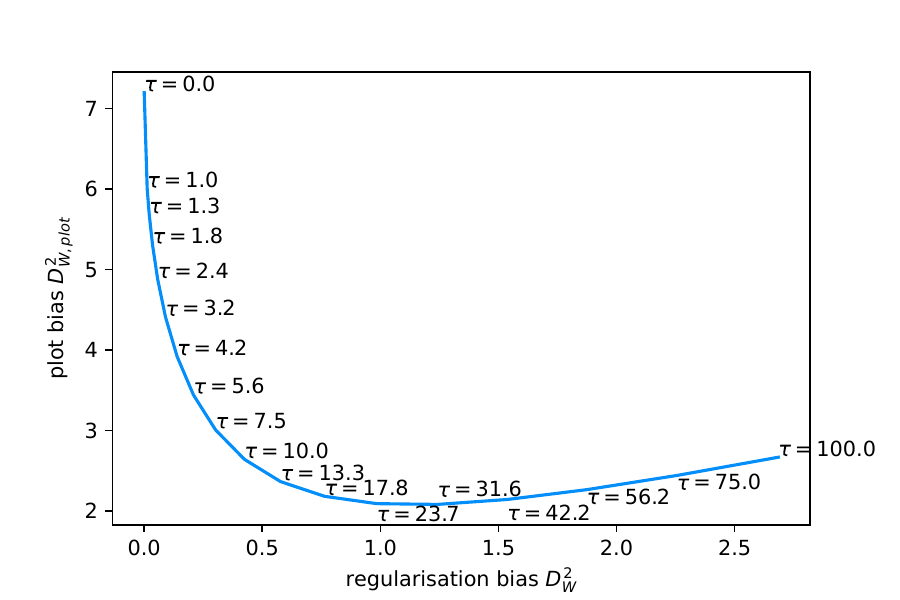}
    \caption{L-curve with plot bias.
    When increasing the regularisation strength, the value of the plot bias (as measured by the W-distance of the regularised result to the unregularised result while ignoring the correlations of the regularised result) decreases up to a certain point, after which it increases again.
    The regularisation bias introduced to the result (as measured by the W-distance of the full regularised result to the unregularised result) increases.
    When viewing the regularisation purely as a data visualisation technique to create less-misleading plots, the optimal choice for the regularisation strength would be at the minimum of the plot bias.
    }
    \label{fig:plot-L-curve}
\end{figure*}

As is illustrated in \autoref{fig:plot-L-curve}, the plot bias will generally have a minimum for a specific value of the regularisation strength.
When viewing the regularisation purely as a data visualisation technique to create less-confusing plots given correlated data-points,
it makes sense to choose the regularisation strength that minimises the plot bias.
The introduction of regularisation bias does not matter in the decision here,
because the fully correct statistical information about the likelihood/posterior distribution is still available from the unregularised result or the combination of the regularised result with the regularisation matrix $A$.
Here we are only concerned about making a plot that reflects this underlying distribution as faithfully as possible.
In the following examples, we will use this method of minimising the plot bias to determine a regularisation strength, unless otherwise specified.

\section{A 2D example}

It's hard to visualise correlations in data with multiple dimensions/bins.
An example with only two bins should illustrate certain features and idiosyncrasies of the regularisation scheme.

Let us assume a result of some measurement with two bins, $\theta_1$ and $\theta_2$. The MLE and its covariance matrix are:
\begin{gather}
    \bm{\hat\theta} = \mqty(\hat\theta_1 \\ \hat\theta_2) = \mqty(5 \\ 3)\text{,} \\
    V = \mqty(\sigma_{\theta_1}^2 & c \sigma_{\theta_1} \sigma_{\theta_2} \\  c \sigma_{\theta_1} \sigma_{\theta_2} & \sigma_{\theta_2}^2) =  \mqty(1.5^2 & \pm 0.95 \cdot 1.5\\ \pm 0.95 \cdot 1.5 & 1) \text{,}
\end{gather}
with two possible values for the correlation coefficient $c = \pm 0.95$.
As the penalty matrix, we choose
\begin{equation}
    Q_1 = \mqty(1 & -1 \\ -1 & 1) \text{,}
\end{equation}
so it penalises differences between the two parameters.
Additionally, let us consider four models that predict certain values for $\bm\theta$ and see what we can deduce about them from comparison to the data.

\begin{figure*}
    \centering
    \begin{subfigure}{0.49\textwidth}
        \includegraphics[width=\textwidth]{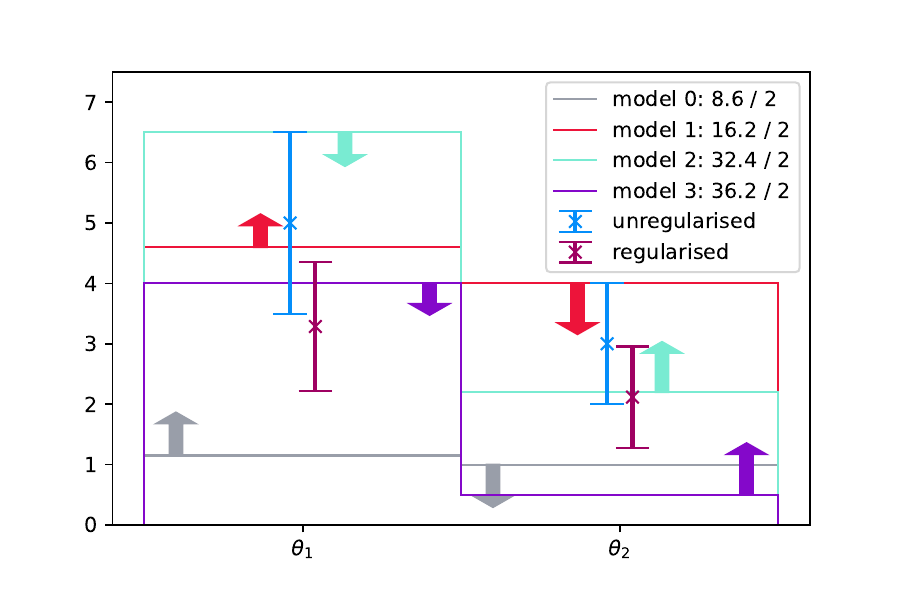}
        \caption{}
        \label{fig:2bins-cor}
    \end{subfigure}
    \hfill
    \begin{subfigure}{0.49\textwidth}
        \includegraphics[width=\textwidth]{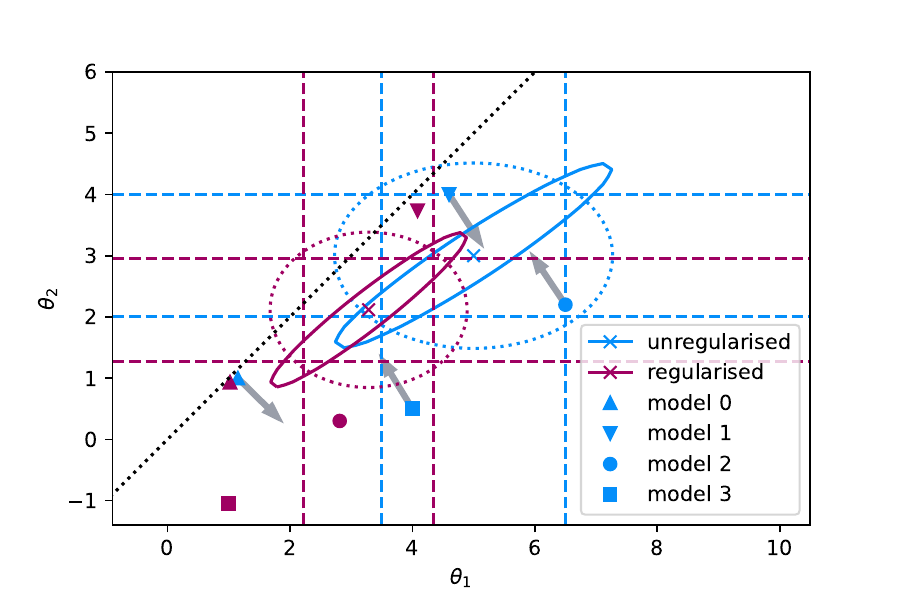}
        \caption{}
        \label{fig:2D-cor}
    \end{subfigure}
    \hfill
    \begin{subfigure}{0.49\textwidth}
        \includegraphics[width=\textwidth]{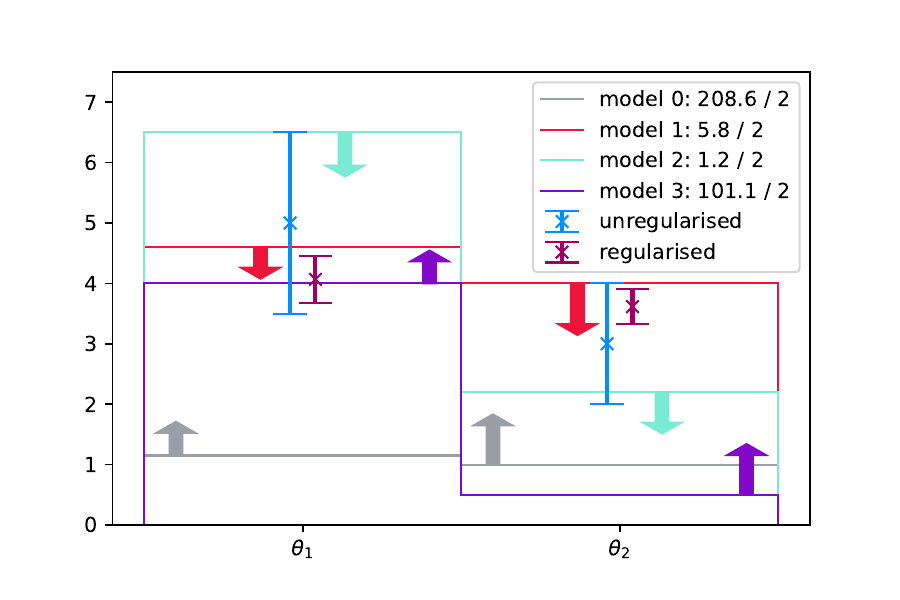}
        \caption{}
        \label{fig:2bins-anticor}
    \end{subfigure}
    \hfill
    \begin{subfigure}{0.49\textwidth}
        \includegraphics[width=\textwidth]{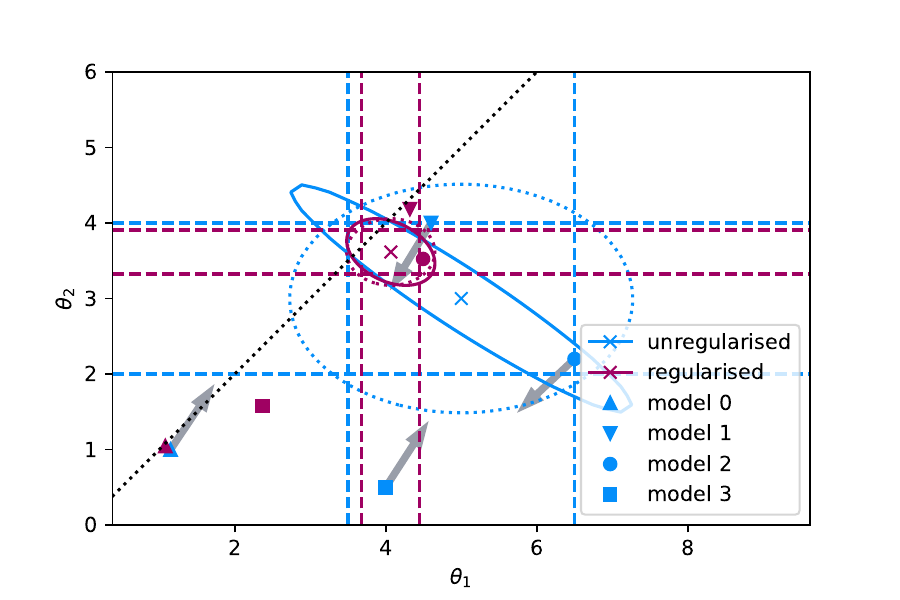}
        \caption{}
        \label{fig:2D-anticor}
    \end{subfigure}
    \caption{Toy example of measurement with two bins with unregularised and regularised data points.
    The upper plots show the case for a positive correlation coefficient of $c=0.95$, and the lower plots for a negative correlation coefficient of $c=-0.95$.
    The arrows on the model predictions show the direction of the local gradient of the unregularised M-distance (gray arrows on the right).
    The numbers in the legend show the squared M-distance of the models to the unregularised data and degrees of freedom of the expected chi-squared distribution.
    The dashed lines in the 2D plot on the right show the extent of the shown uncertainties of the single parameters on the left.
    The dotted contours show the uncorrelated $68\%$ confidence region implied by those uncertainties.
    The solid contours show the actual confidence region when considering the correlation of the parameters.
    The diagonal dotted line shows the regularisation target, where $\theta_1 = \theta_2$.}
    \label{fig:2bins}
\end{figure*}

\autoref{fig:2bins-cor} shows the positively correlated data as it would be presented if it had more dimensions/bins.
Each parameter is shown with its MLE and the uncertainty corresponding to the diagonal element of the covariance matrix.
Because the uncertainties are actually correlated, it is important to also show the goodness of fit of the models, i.e. the squared M-distance to the unregularised data.
Otherwise it would not be possible to deduce, e.g. that model~0 actually fits the data better than model~1 in the positively correlated case.

Even with the M-distance displayed, it is not clear why model~0 should fit the data better than model 1 or~2.
It does become obvious when looking at the situation in the 2D plot (\autoref{fig:2D-cor}), though.
Here it is possible to show the effect of the correlation in the shape of the confidence region.
These ellipses are lines of constant M-distance.
It is clear that models 1 and~2 are more ``sigmas'' away from the data than model~0.

This seemingly ``wrong'' behaviour of preferring a model that looks ``further away'' from the data points in a histogram is known as ``Peelle's Pertinent Puzzle'' (PPP) in the context of fitting models to correlated data \cite{Chiba1991, Smith2007}.
In fact, model~0 and model~1 have the same ratio of $\theta_1 / \theta_2$.
So if one were to fit a scaling factor of these models to the data, the fit would prefer a result closer to model~0.
The correlation also explains why models 2~and~3 actually fit the data worse than what the comparison with the unregularised histogram suggests.

It is impossible to show the complete correlation information in an N-dimensional histogram plot.
But it is possible to show at least some additional information about the likelihood/M-distance surface around the model parameter points.
By adding the components of the local gradient (gray arrows in \autoref{fig:2D-cor}) to the bins, it becomes clear that the position relative to the MLE point of the data is not the whole picture.
It is still not obvious that model~0 should be preferred over model~1,
but it shows that even at the point of model~0, there seems to be stronger tension in the ratio rather than the normalisation of the parameters.

Now let us see how the regularisation affects things.
The penalty term moves the MLE closer to the $\theta_1 = \theta_2$ diagonal.
Because of the correlation of the shape and normalisation, the MLE also moves to a lower overall normalisation of $\bm\theta$, despite the penalty term only affecting the shape.
Note that this kind of effect can be present independently of how one chooses the regularisation strength, and even if the uncertainties on $\theta_1$ and $\theta_2$ are uncorrelated.
The shape and norm can be correlated simply by the fact that they are functions of the same parameters.

In the scheme presented in this paper, the regularisation strength is chosen to minimize the difference between the implied uncorrelated distribution of the regularised result (the dotted burgundy coloured contour) and the actual correlated distrbution of the unregularised result (solid blue contour in \autoref{fig:2D-cor}).
As expected the overall size of the shown uncertainties is reduced.
But of course this uncertainty now covers only a subset of the original confidence region and is biased towards the regularisation target.

Despite the regularisation, it is still not possible to understand the full picture by looking at the histograms alone.
The local gradient at model~3 agrees with the naive conclusion when just comparing the position of the regularised MLE with the model.
For the other models, the local gradient pulls the parameters to a different shape compared to the data though.
The regularisation does not replace a proper GOF calculation, and additional information about the gradient is still helpful in understanding the tension between data and model.

The shown models are \emph{not} regularised in the histogram plots, i.e. they are \emph{not} folded through the regularisation matrix $A$ when plotting them in the histogram, even when comparing them to the regularised data.
Calculating the M-distance always happens with the unregularised data and covariance, or equivalently by multiplying the models with $A$ and using the regularised data and covariance.
This is because in these plots, we keep the meaning of the bin values the same, even when regularising the data by multiplying it with $A$.
And for that meaning of the bin values, the original model should be shown.
The regularisation strength is also chosen to make the shown uncorrelated distribution as close as possible to the real correlated distribution, implying an unchanged meaning of the bin values.

Additionally, trying to regularise the models can lead to unexpected effects.
If a model is less ``regular'' than the data itself, i.e. if it gets a worse penalty term than the data, it gets moved around more than the data.
For example, model~3 is moved to negative values of $\theta_2$ when multiplied with $A$ (burgundy square in \autoref{fig:2D-cor}).
Depending on the meaning of the parameter, this could make the model prediction unphysical.
Of course the axis offsets and scaling are arbitrary in this example,
but it should serve as an illustration of what could happen.

Let us now take a look at the case with anticorrelated data (plots of \autoref{fig:2bins-anticor} and \ref{fig:2D-anticor}).
Compared to the correlated case, the unregularised data looks exactly the same in the histogram plot.
But the GOF scores have changed dramatically because of the different correlation, again showing that looking at a histogram alone is unsuitable to judge the GOF of a model to correlated data.

Because of the orientation of the correlation, the regularisation did not pull the norm of $\bm\theta$ in unexpected directions this time.
The resulting uncertainties of the regularised data are quite a bit smaller compared to the previous case, and the regularised covariance matrix has very little correlation.
The local gradients and GOF scores of models 0~and~1 now agree with the naive comparison with the regularised data in the histogram.
Models 2~and~3 on the other hand still show a local pull to a different shape compared to the regularised result.
Model~2, even though it looks like it should be in strong tension with the regularised data, is actually a very good fit, when considering the GOF score.
All of this illustrates again that it is difficult to get a complete picture from just the histogram or just the GOF score.
They should be seen in combination and potentially augmented by the local gradient information.
Note also that the ``regularised'' models are in different positions now.
This reflects the fact that $A$ is determined by the interplay of regularisation condition and actual covariance in the data.

\section{A real-world example}

We can test the approach by applying it to a real cross-section measurement.
We choose the $\delta p_T$ measurement of \cite{T2K:2018rnz}.
It is a simple single-differential measurement, both unregularised and regularised results are vailable in a data release \cite{T2K:2018rnzdata}, and the original author provided us with the original Monte Carlo predictions that were used as a basis for the unfolding technique.
This allows us to make the post-hoc regularisation as close as possible to the originally employed one.
Differences remain though, and we cannot expect the results to be identical.

This measurement was unfolded by doing a template parameter fit.
The nominal MC model was divided into the true analysis bins and each bin corresponds to a template in the expected reconstructed event distributions.
The fitter was then free to vary the weights of each truth bin, as well as additional nuisance parameters, to maximise the likelihood of the reco space expectation given the actual observed data.
The resulting MLE of the template weights and nuisance parameters as well as the correlated uncertainties are then propagated to the final cross-section result.

For the regularised result, the authors add a penalty term to the likelihood function which is proportional to the sum of the squared differences of neighbouring template weights.
These weights are all 1 for the nominal input model, so this is equivalent to using the $Q_{1m}$ matrix described above.
The authors used the L-curve method to determine the regularisation strength.

We cannot replicate this regularisation exactly with the post-hoc approach, since we only have access the the final cross-section result, and not the intermediary step of the template weight and nuisance parameters.
Our additional penalty term will thus not be exactly the same as the original one, but by using the $Q_{1m}$ matrix with the provided nominal model, we can come pretty close.

\begin{figure*}
    \centering
    \includegraphics{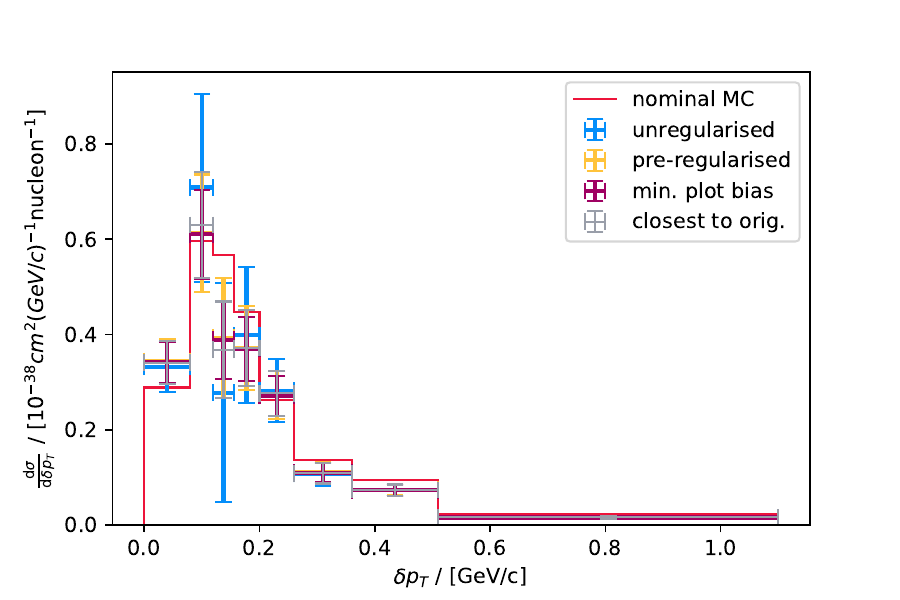}
    \caption{Transverse momentum imbalance measurement with different regularisations.
    The unregularised and pre-regularised data points are taken from \cite{T2K:2018rnzdata}.
    The nominal model was provided by the original authors of \cite{T2K:2018rnz}.
    The other two sets of points were created with the post-hoc regularisation approach and a model scaling penalty matrix $Q_{1m}$ using the nominal model.
    The regularisation strength was chosen to minimise the plot bias $D^2_\text{W,plot}$ and the Wasserstein distance to the pre-regularised result $D^2_\text{W,pre-reg}$ respectively.
    The post-hoc regularisation results are very close to the original one, with $D^2_\text{W,pre-reg} = 0.35$ and $D^2_\text{W,pre-reg} = 0.14$.
    }
    \label{fig:dpT}
\end{figure*}

We can measure how close the post-hoc regularisation comes to the original one by calculating the W-distance $D^2_\text{W,pre-reg}$ between the two.
In this case we use the original regularsied covariance matrix as basis for the whitening transformation.
\autoref{fig:dpT} shows the original results as well as the post-hoc regularisation.
The regularisation strength $\tau$ was chosen in two different ways:
once by minimising the plot bias $D^2_\text{W, plot}$ as you would do with any unregularised data set,
and once by minimising $D^2_\text{W,pre-reg}$.
The former gives an idea of how different the regularised result is when using the regularisation scheme suggested in this paper.
The latter can tell us how much of that difference comes from the difference in penalty terms alone,
ignoring the different ways of choosing the regularisation strength.
I.e., if the penalty term was identical, there should be a regularisation strength that reduces the difference to the pre-regularised result to 0.
The actual W-distances to the pre-regularised result are 0.35 and 0.14 respectively.
This should be compared to the number of bins, 8, as it is a generalisation of the M-distance.
So there is a difference in the regularisation as expected, but it is small.

\begin{figure*}
    \centering
    \includegraphics{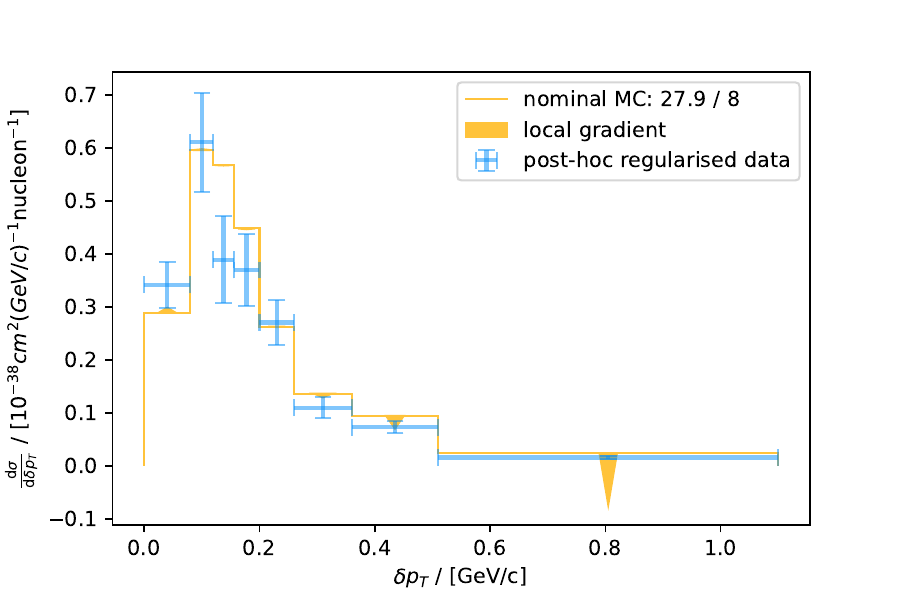}
    \caption{Local M-distance gradient of the nominal model.
    The M-distance is calculated with the unregularised result.
    The gradient is dominated by the bin with the smallest value and similarly small associated uncertainty.
    The length of the gradient is scaled so the largest component has an absolute value of $\SI{0.1e-38}{cm^2 GeV^{-1} c\,nucleon^{-1}}$}.
    \label{fig:dpT-grad}
\end{figure*}

Let us now see what we can learn from looking at the local M-distance gradient at the model.
\autoref{fig:dpT-grad} shows the nominal model together with the post-hoc regularised data and the local M-distance gradient at the model prediction.
The length of the gradient arrows was scaled arbitrarily to give the biggest component a length of $\SI{0.1e-38}{cm^2 nucleon^{-1} GeV^{-1}c}$.
It seems like the fastest way to imrpove the model locally, is to reduce the cross section in the last bin.
But this is somewhat misleading, since the gradient is calculated in terms of the absolute values of the bins.
The last bin's small model expectation and also small (in absolute terms) uncertainty, mean that the M-distance changes a lot compared to the same absolute change in differential cross section in the other bins.

\begin{figure*}
    \centering
    \includegraphics{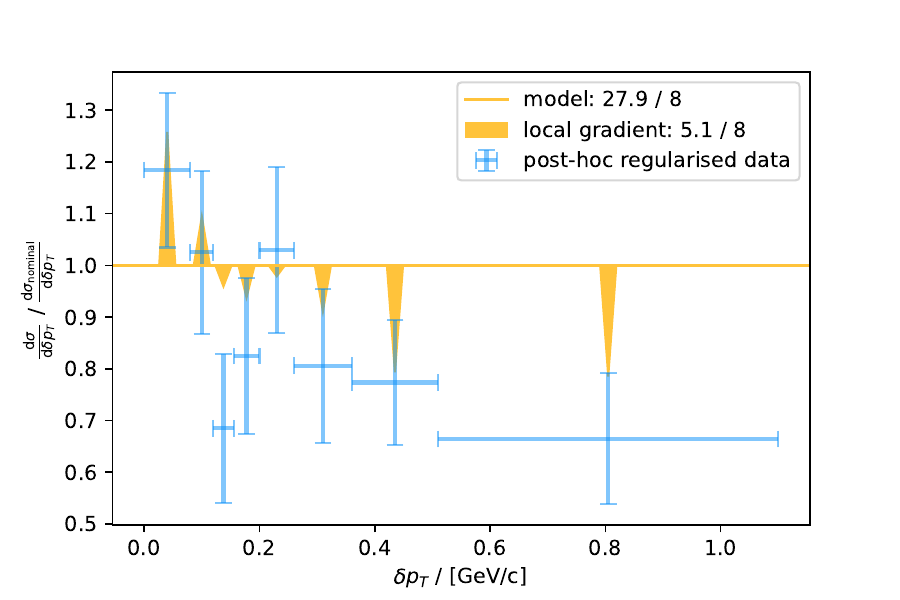}
    \caption{Local M-distance gradient of the nominal model in a cross-section ratio plot.
    The M-distance is calculated with the unregularised result.
    This parameterisation is a linear transformation of the original result, so the M-distance of the model is exactly the same as in the original as shown in \autoref{fig:dpT-grad}.
    The gradient in this parameterisation is no longer dominated by a single bin.
    The gradient is scaled so that its endpoint has the lowest possible M-distance.
    At an M-distance of 5.1, this endpoint is much more compatible with the data (p-value: 0.75) than the nominal model (p-value: 0.0005).
    Note that the local gradient does not not lower the cross section around \SI{0.2}{GeV/c} by as much as the data points would suggest.}
    \label{fig:dpT-ratio}
\end{figure*}

This shows that while the gradient adds additional information to the plot, one needs to think carefully about what question it actually answers.
In this case, it tells us how to improve the model the quickest locally, in terms of adding or removing differential cross section.
But this might not be the most useful way of presenting the data.
One often is interested in the data to Monte Carlo ratio, exactly because it avoids some problems with looking at bins with very different values for the cross section.
\autoref{fig:dpT-ratio} shows the same data and model as \autoref{fig:dpT-grad}, but now all data points are normalised to the nominal model.

The gradient in this normalisation factor is different from the gradient in the differential cross-section space.
It is \emph{not} just the differential cross-section gradient divided by the model.
In fact, it is actually multiplied by the model values:
\begin{gather}
    \sigma_i = \eval{\dv{\sigma}{\delta p_T}}_i \text{,} \\
    \nabla{}_{\sigma,i} = \dv{D^2_M}{\sigma_i} \\
    x_i = \eval{\dv{\sigma}{\delta p_T}}_i\ /\ \eval{\dv{\sigma_\mathrm{nominal}}{\delta p_T}}_i \text{,} \\
    \nabla{}_{x,i} = \dv{D^2_M}{x_i} =
    \nabla{}_{\sigma,i} / \dv{x_i}{\sigma_i} = \nabla{}_{\sigma,i} \eval{\dv{\sigma_\mathrm{nominal}}{\delta p_T}}_i \text{.}
\end{gather}
Note that the nominal model values here are used as a constant scaling parameter.
They are \emph{not} considered as a variable.
This means that presenting the cross section as a ratio to the model is a linear transformation and all information about the likelihood is perfectly preserved.
So this transformation can be done without modifying the information of the result in any way, which makes it ``safe'' to do with other people's data.
This is in contrast to other possible ways of displaying data with bin values that span multiple orders of magnitude, like e.g. a presentation in log-space.

The gradient in \autoref{fig:dpT-ratio} is no longer dominated by the last bin.
The length of the arrows is now chosen to point to the minimum achievable M-distance along the direction of the local gradient.
The nominal model is incompatible with the data at an M-distance of 27.9 and 8 degrees of freedom, which corresponds to a p-value of 0.0005.
The best fit along the gradient is perfectly compatible with the data at an M-distance of 5.1, corresponding to a p-value of 0.75.
Note that the local gradient does lower the cross-section in the $\SI{0.2}{GeV/c}$ region, but not as much as the data points would suggest.
The minimum achievable distance overall would of course be a model that agrees perfectly with the unregularised data points.

\section{conclusions}

Unregulraised, unfolded cross sections provide statistically ``correct'' information and are the least biased, but correlations make interpreting plotted histograms difficult.
Regularisation can reduce correlations and variation at the cost of introducing some bias.
To get the best of both worlds, one can provide unregularised results to use with statistical tests and global fits, as well as regularised results for the presentation in plots that are easier to interpret.
This can lead to the question which of these two is the ``real'' result of an analysis.
Understanding regularisation as linear transformation of the unregularised results allows us to draw a direct connection between the two.

Providing the regularised result together with the regularisation matrix $A$, allows the same unbiased calculation of M-distance and p-values as the unregularised result.
No information is lost, since the matrix $A$ can be seen as a simple linear coordinate transformation.
The models just need to be folded through the $A$ matrix when doing any quantitative evaluation.
This means it is not necessary to provide two separate results, avoiding any potential confusion.

When plotting the data and comparing it to models qualitatively, the models should \emph{not} be folded through $A$, though.
Instead it is better to show them unmodified.
This is because the regularisation matrix can modify the model prediction in ways that make drawing correct conclusions from the plot impossible.
This is especially likely if the models themselves would receive a stronger regularisation penalty term than the unregularised data,
i.e. if they are ``further away'' from the regularisation target than the data itself.
Conceptually, when plotting, the $A$ matrix should be viewed as modifying the result, rather than the coordinate system,
since the axes', i.e. the bin contents', meaning is given by the plot.
Since the axes have their original meaning, the models should also be shown unmodified.

Comparing the regularised result with the original model shows one possible way to improve the fit of the model to the data: modifying the model so it looks more like the regularised result.
The regularisation is used to pick some subset of the possible result space that is compatible with the data, using the regularisation penalty term as tiebreaker.
Other equally valid solutions are possible though.
The final conclusion of whether a particular change of a model improves or worsens the agreement with the data should always be drawn using the unregularised M-distance, or equivalently the M-distance calculated using the regularisation matrix $A$.
Showing the local gradient in unregularised M-distance at the point of the model prediction can provide additional information about where the model is ``pulled'' by the data, including effects from correlations.

A distribution plotted as histogram with central values and error bars is always implicitly uncorrelated, since it is impossible to display the correlations between the data points.
The W-distance $D^2_\text{W,plot}$ can be used to measure the difference between the shown, uncorrelated distribution with the actual, correlated, unregularised distribution.
This ``plot bias'' can be used as an objective loss function for the optimisation of regularisation strengths of Tikhonov regularisation.
Rather than trying to optimise for some statistical property of the regularised result,
this acknowledges that the regularisation is mostly a technique for data presentation purposes.
The correct statistical properties are retained by using the regularisation matrix $A$ on the tested models.

Using the linear algebra approach presented in this paper, it is possible to apply Tikhonov regularisation to any binned spectrum that is presented as a central value and covariance matrix.
No detailed knowledge about the genesis of the result is necessary.
The only assumption made is that the provided central value and covariance matrix are a suitable approximation of the likelihood function in the case of Frequentist analyses, or the posterior probability in the Bayesian case.
This allows the regularisation of results after they have been published,
and could thus be called ``post-hoc regularisation''.

Condensing everything down to a set of suggestions:
\begin{itemize}
    \item Unfold without any regularisation, yielding an unregularised result $\bm{\hat\theta}$ and $V$
    \item Use method described in this paper or from \cite{Tang2017} to create a regularisation matrix $A$.
    \item Use regularised result $\bm{\hat\theta}' = A\bm{\hat\theta}$ and $V' = AVA^T$ for plots.
    \item Do \emph{not} multiply model predictions $\bm{m}$ with $A$ when plotting comparisons.
    \item Use the unregularised M-distance $D_M^2 = (\bm{m} - \bm{\hat\theta})^TV^{-1}(\bm{m} - \bm{\hat\theta}) = (A\bm{m} - \bm{\hat\theta}')^TV'^{-1}(A\bm{m} - \bm{\hat\theta}')$ to calculate GOF scores.
    \item Always provide GOF scores in plots.
    \item Supplement plots by adding information about the local GOF gradient of model predictions.
\end{itemize}
Following these suggestions does not eliminate the challenges of presenting and discussing strongly correlated data sets, but it should help in the effort to be as little misleading as possible.

\section{Acknowledgements}

I would like to thank Steven Gardiner, Elena Gramellini, Krishan Mistry, Xin Qian, Hanyu Wei, Wenqiang Gu, and Andy Furmanski (in no particular order) for some very useful discussions about the Wiener-SVD method and their application in the MicroBooNE experiment.
I also would like to thank Kirsty Duffy for facilitating these discussions.
Further thanks go to Stephen Dolan for providing the original model predictions used in T2K's $\delta p_T$ analysis.
This work was supported by a grant from the Science and Technology Facilities Council.
It was funded by the Deutsche Forschungsgemeinschaft (DFG, German Research Foundation) under Germany’s Excellence Strategy – EXC 2118 PRISMA+ – 390831469.

\bibliography{biblio}%

\end{document}